# Physical Probability and Locality in No-Collapse Quantum Theory[1]

Simon Saunders [2]

**Abstract.** Probability is distinguished into two kinds: physical and epistemic, also, but less accurately, called objective and subjective. Simple postulates are given for physical probability, the only novel one being a locality condition. Translated into no-collapse quantum mechanics, without hidden variables, the postulates imply that the elements in any equiamplitude expansion of the quantum state are equiprobable. Such expansions therefore provide ensembles of microstates that can be used to define probabilities in the manner of frequentism, in von Mises' sense (where the probability of $P$ is the frequency of occurrence of $P$ in a suitable ensemble). The result is the Born rule. Since satisfying our postulates, and in particular the locality condition (meaning no action-at-a-distance), these probabilities for no-collapse quantum mechanics are perfectly local, even though they violate Bell inequalities. The latter can be traced to a violation of outcome independence, used to derive the inequalities. But in no-collapse theory that is not a locality condition; it is a criterion for entanglement, not locality.

## 1. Introduction

By physical probability (also called 'objective probability' or 'chance'), I mean probability as something 'out there' in the world, independent of human agency and rational choice theory, and patterns of reasoning more generally. The contrast is with epistemic probability (also called 'subjective probability', or 'credence'), that does depend on those things.

Although long hinted at, for example, by Laplace, the clear recognition of the distinction is relatively recent, in Mellor (1971), Hacking (1975), and especially Lewis (1980). It has had a large impact in philosophy; in physics, not so much. Yet ambiguity on the meaning of probability can seldom have been more damaging than in the case of the Born rule of quantum mechanics. If it is a rule for physical probability, it is not obvious it must involve the notion of uncertainty, clearly an epistemic notion; whereas if it is a rule for epistemic probability, of what, rationally, to believe, then it is not up to physics to say (as in 'Postulate 1: believe the following postulates').

Here we consider only physical probability. We offer four postulates, three of which are widely if not universally assumed, at least in physics. The fourth is different. It is clearly close to a locality principle -- in that it turns into one, when implemented in a theory in which everything other than probability is local. In that case, and looking ahead, when applied to the set-up envisaged by Einstein, Podolsky, and Rosen, as reformulated by Bohm (the EPRB set-up), it clearly is a locality condition. By 'locality', unless otherwise qualified, I mean no action-at-a-distance.

---

We translate those postulates into quantum mechanics, and specifically no-collapse quantum theory, where the dynamics is always unitary. This is, from a realist point of view, a many-worlds theory, but the details need not concern us, for all we need is no-collapse, unitary dynamics, and no hidden-variables.

It then follows that states entering into a superposition, yielding the total state, if assigned probabilities, *must be assigned equal probabilities if their amplitudes (norms) are the same.* The result holds with complete universality. But then, if the quantum state is expanded in terms of equiamplitude eigenstates of $P$ (some projector operator), the fraction in the expansion, with eigenvalue +1 of $P$, is the probability of $P$. Any other verdict would violate our postulates.

This conception of probability is familiar from the writings of von Mises and others, where it is known as 'frequentism': probability is frequency of occurrence in a given ensemble. 'Fractionalism' would be the better term for us, for as we shall see, it may be only a superposition of states has some property $P$, and not states taken singly. We shall call it 'microstate-counting', which accurately conveys the idea.

Our main result is that the probability of $P$, thus construed, in the limit when the number of states in the equiamplitude expansions increases without limit, is the Born rule quantity for $P$. The result is of interest in its own right, because unlike Gleason's theorem, we make no assumption of noncontextuality. But it further sets up an interesting dilemma, as follows.

Since implying the Born rule, probabilities understood in this way must also, in some circumstances, violate Bell inequalities, despite satisfying our locality postulate. Therefore, they must violate either parameter independence or conditional outcome independence, or both, the two conditions which, jointly satisfied, imply factorizability, and hence Bell inequalities (see e.g. Myrvold et al (2023) for terminology and the notation we use later). Which, or both?

The answer is that our locality postulate, in the Bell set-up, *just is* parameter independence. The postulate would also imply conditional outcome independence, and hence factorizability, but only if experiments have unique outcomes. In no-collapse theory, a superposition of outcomes is produced by remote experiments, and locality with respect to that is just to restate parameter independence. There is no further condition in no-collapse theory, based on locality, in the sense of our postulates.

That does not mean outcome independence has no meaning in no-collapse theory; on the contrary, it is a criterion for separability. As such, in the EPRB set-up, where the two spin systems are in an entangled state, it is rightly violated.

That sets up the dilemma. As argued in Maudlin (2011, 2014), and endorsed by many others, but making explicit the assumption that experiments have unique outcomes (ONE), and assuming no retrocausation and the like, the meaning of the observed violations of Bell inequalities (BV) is

$$ONE \wedge BV \rightarrow \neg LOC$$

where $LOC$ is locality, independent of any special form of realism. Taken to its logical conclusion, as by Valentini (2025) in pilot-wave theory, the implications for physics are revolutionary. But this sentence is logically equivalent to

$$BV \wedge LOC \rightarrow \neg ONE.$$

The latter seems the more appropriate way of stating the importance of BV, as there is abundant evidence that LOC is a fundamental principle of nature; whereas there is no evidence at all that there is only a single world. It is revolutionary in a different way. (A similar claim is made in Waegell and McQueen (2020), but on different grounds; see Faglia (2024) for discussion.)

## 2. Postulates for Physical Probability

P1  Boolean event space: there exists a Boolean space $\Sigma$ of elements $X \in \Sigma$ equipped with disjointness, union, and complement.

P2  Instantaneous state: the physical probability $\mu$ of $X$ depends only on the instantaneous physical state and $\Sigma$.

P3  Formal: $\mu[X]$ is bounded by 0 and 1, $\mu[\Sigma] = 1$, $\mu[\emptyset] = 0$; if $X, Y$ disjoint, then $\mu[X \cup Y] = \mu[X] + \mu[Y]$

P4  Locality: $\mu[X]$ can only change if there is a physical change in $X$.

The space $\Sigma$ is intended to be as general as possible, so $X, Y$ etc may be things, states, events, or properties – call 'ontology' – at a given time (to fit with P2). In order that Postulate 4 be non-trivial, the physical change in $X$ cannot consist only in the change in its physical probability. The space $\Sigma$, the physical state, the notion of physical change -- all these are to be specified by candidate physical theories.

In justification (and here I avoid comment on quantum mechanics, as I seek generality): Postulate 1 appears innocuous; some such notion appears inevitable. Postulate 2 was used in Einstein's historic 1905 paper, where it was called 'Boltzmann's principle': the entropy of a system is a function of the probability of its instantaneous state. Using this, he compared the entropy change in an ideal gas, with that in black-body radiation, due to fluctuation, in the Wien regime, to conclude they were the same – this his main argument for light quanta.

Postulate 3 is standard. It is Postulate 4 that is new. Suppose it is realised in a physical theory in which $X$ and $Y$ are spatially remote (naturally our postulates say nothing about any spatial geometry); then it says the physical probability of $X$ cannot be changed by a change in $Y$ alone, including a change in probabilities at $Y$. It does not forbid non-locality – the theory may be non-local, and Postulate 4 still be satisfied (through action-at-a-distance by $Y$ on $X$). But in a physical theory in which things, the ontology, only interact locally ('local causality', as Bell called it), Postulate 4 says that physical probability must be local as well. Evidently, in light of P2, the physical state and the ontology have to be closely connected.

Postulate 4 has another virtue: it highlights the difference between physical and epistemic probability. Evidently P1 poses no problem for credence, nor P3, which follows from Dutch Book arguments; and perhaps it can be straightjacketed into P2; it is P4 that seems strange. Suppose two envelopes are sent, one to Alice and one to Bob, each remote from the other. One envelope is empty, the other contains a $5 bill. You don't know which envelope was sent to which place; you give it even odds.  Alice opens her envelope and finds the $5 bill. The probability that it contains the bill is changed to 1 – and the probability that *Bob's* envelope, however remote, contains the bill, changes to 0, although nothing else in Bob's location is physically changed. Postulate 4 is clearly violated. There seems no reason, on epistemic grounds, to subscribe to this principle.  Yet if probability is something physical, and applicable to events causally separated from one another – to a theory in which causal action is otherwise local – it is another story. Arguably, probability should then be local as well. If there is such a thing as physical probability, in a local physical theory, it should not involve any spooky action-at-a-distance.

We proceed by translating P1-P4 into quantum mechanics, where the physical state is the total quantum state, an element of a Hilbert space $\mathcal{H}$, and physical changes are unitary transformations on $\mathcal{H}$, for which a Schrödinger equation holds, and the superposition principle holds without restriction.

## 3. Expansions of the state as event spaces

By 'state' we always mean a vector state in $\mathcal{H}$, complete with (real) amplitude and (complex) phase, with no assumption it is normalised (so we use the notation $\psi$ rather than $|\psi\rangle$). We take as the physical state some vector $\psi \in \mathcal{H}$.

Consider an arbitrary expansion of $\psi$ in finitely many orthogonal vectors

$$\psi = \varphi_1 + \ldots + \varphi_n. \tag{1}$$

We take as the space $\Sigma$ the vectors in Eq.(1), and (vector) sums of such, together with the zero vector. Disjointness is given by orthogonality, and the complement, for any $\varphi \in \Sigma$, is $(I - P_\varphi)\psi \in \Sigma$. (This is easily turned into a Boolean event space, treating superpositions of vectors as sets of vectors, with meet and join given by set-theoretic intersection and union.)

This may look unfamiliar; more standard would be to define a Boolean algebra in terms of some (commuting) family of projectors $P$; but any projector $P$ so defined may be mapped to the vector $P\psi$, returning us to the vector space. (The ontology consists of vectors in $\mathcal{H}$, not projectors.)

We assume probabilities are real-number assignments depending only on $\psi, \Sigma$, so we write the wanted probability distribution as $\mu_{\psi_\Sigma}: \varphi \in \Sigma \rightarrow \mu_{\psi_\Sigma}[\varphi] \in [0,1] \subset \mathbb{R}$. In line with P3, we assume pairwise additivity:

$$\varphi, \eta \in \Sigma, \langle \varphi | \eta \rangle = 0 \Rightarrow \mu_{\psi_\Sigma}[\varphi + \eta] = \mu_{\psi_\Sigma}[\varphi] + \mu_{\psi_\Sigma}[\eta]. \tag{2}$$

It follows from Eq.(2) and the definition of the complement that $\mu_{\psi_\Sigma}[\psi] = 1$, $\mu_{\psi_\Sigma}[\emptyset] = 0$. Notice that P1-P3, translated into quantum mechanics, do not imply Gleason's theorem, as there is nothing in these postulates requiring that a probability measure be defined over non-commuting projectors, as Gleason assumed. P1 only requires the existence of a Boolean event space.

In implementing Postulate 4, we assume that physical changes involve only unitary mappings $U$. Contraposing, it says if $\varphi \in \Sigma$ is unchanged under $U$, its probability is unchanged, that is:

$$\varphi \in \Sigma, \ U\varphi = \varphi \Rightarrow \mu_{U\psi_\Sigma}[\varphi] = \mu_{\psi_\Sigma}[\varphi] \tag{3}$$

where $U\psi_\Sigma$ is the state $U\psi$ with event space $U\Sigma$ generated by $U\varphi_1, \ldots, U\varphi_n$. Notice that if $U\varphi = \varphi$, then $\varphi \in U\Sigma$, so the probability $\mu_{U\psi_\Sigma}[\varphi]$ is well-defined.

This completes the translation of our postulates into no-collapse quantum mechanics.

Given Eqs.(1)-(3), it is now easy to show that if a pair of states in $\Sigma$ have the same amplitude, then they have the same probability. For convenience, we write Eq.(1) in the notation:

$$\psi = \varphi_a + \varphi_b + \varphi_1 + \ldots + \varphi_{n-2}. \tag{4}$$

Our proof is similar to that in Short (2023), save that, importantly, ours makes use of only unitary transformations. Let $U_a$ be a unitary transformation that acts as the identity on $\varphi_b$ and all $\varphi_k$, with the action $U_a: \varphi_a \rightarrow \varphi_c$, where $\varphi_c$ is orthogonal to all the states on the RHS of (4). Similarly, let $U_b$ act as the identity on $\varphi_c$ and all $\varphi_k$, with the action $U_b: \varphi_b \rightarrow z_b \varphi_a$, for some $z_b \in \mathbb{C}$; and finally, let $U_c$ act as the identity on $\varphi_a$ and all $\varphi_k$, with the action $U_c: \varphi_c \rightarrow z_a \varphi_b$, for some $z_a \in \mathbb{C}$.

Let $\mu_{\psi_\Sigma}$ be any assignment of probabilities to $\varphi \in \Sigma$ satisfying (2),(3) and (4). From (3), noting that the antecedent is satisfied when $U$ is $U_a$ and $\varphi$ is $\varphi_b$ or one of the $\varphi_k$'s:

$$\mu_{U_a\psi_\Sigma}[\varphi_b] = \mu_{\psi_\Sigma}[\varphi_b]; \ \mu_{U_a\psi_\Sigma}[\varphi_k] = \mu_{\psi_\Sigma}[\varphi_k], \ k = 1, \ldots, n-2.$$

From additivity, and since $\mu_{U_a\psi_\Sigma}[U_a\psi] = 1$, the sum of these numbers with $\mu_{U_a\psi_\Sigma}[\varphi_c]$ is unity, it follows:

$$\mu_{U_a \psi_\Sigma}[\varphi_c] = \mu_{\psi_\Sigma}[\varphi_a].$$

By similar reasoning

$$\mu_{U_b U_a \psi_\Sigma}[\varphi_a] = \mu_{U_a \psi_\Sigma}[\varphi_b] = \mu_{\psi_\Sigma}[\varphi_b]$$

and

$$\mu_{U_c U_b U_a \psi_\Sigma}[\varphi_b] = \mu_{U_b U_a \psi_\Sigma}[\varphi_c] = \mu_{U_a \psi_\Sigma}[\varphi_c] = \mu_{\psi_\Sigma}[\varphi_a]. \tag{5}$$

The result of the three changes is that the probabilities of $\varphi_a$ and $\varphi_b$ have been switched, with all other probabilities unchanged. That seems appropriate: just those two states were physically changed.

The final physical state is:

$$U_c U_b U_a \psi = z_a \varphi_b + z_b \varphi_a + \varphi_1 + \cdots + \varphi_{n-2}.$$

But now it follows that when $\varphi_a$, $\varphi_b$ have the same amplitude, the complex numbers $z_a$, $z_b$ are pure phases; for from unitarity we have

$$\|z_a \varphi_b\| = \|\varphi_c\| = \|\varphi_a\|$$

$$\|z_b \varphi_a\| = \|\varphi_b\|.$$

The phases $z_a, z_b$ can be absorbed into $U_b, U_a$, whereupon $U_c U_b U_a \psi$ and $\psi$ are identical. From Eq.(5) we obtain

$$\mu_{U_c U_b U_a \psi_\Sigma}[\varphi_b] = \mu_{\psi_\Sigma}[\varphi_b] = \mu_{\psi_\Sigma}[\varphi_a]$$

as was to be proved. See Fig.1.

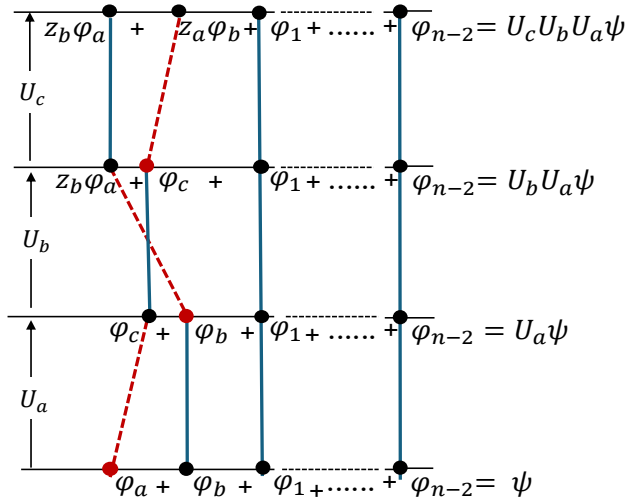

**Figure 1**: Each unitary transformation $U_a$, $U_b$, $U_c$ acts on only one state (shown in red) in the superposition. Since the physical probabilities of the other $n-1$ projectors cannot change, and must sum to one, it follows $\mu_{U_a \psi_\Sigma}[\varphi_c] = \mu_{\psi_\Sigma}[\varphi_a]$. A similar argument applies to $U_b$ and $U_c$.

### 4. Probabilities as fractions of ensembles

In this section we summarise the microstate-counting approach in Saunders (2024), save that now it is freed from any dependence on decoherence theory. However, we assume the dimension of the Hilbert space $\mathcal{H}$ to be infinite, as is required in realistic applications of quantum mechanics (for example, to define physical quantities that may take on continuous values).

The key idea is to expand $\psi$ in equiamplitude orthogonal states, of the form:

$$\psi = \xi_1 + \ldots + \xi_n; \quad \|\xi_k\| = \|\xi_j\|, k, j = 1, \ldots, n \qquad (6)$$

and use these expansions as ensembles, in the definition of probabilities as frequencies -- or better *fractions*. Call the vectors $\xi_k$ *microstates*, and denote such an expansion $\Lambda_\psi^n$.

Such expansions can always be found for any $\psi$, for any $n$, and if one such, uncountably infinitely many (this is true even when $\mathcal{H}$ is finite dimensional, so long as $\dim \mathcal{H} > 2$). If this is not obvious, suppose that an equiamplitude expansion $\Lambda_\psi^m$ of $\psi$ exists for $m < \dim \mathcal{H}$, and consider a vector $\phi$ orthogonal to all the vectors in $\Sigma_\psi^m$ (and hence to $\psi$), with the same norm as the $\xi_k$'s. Rotate $\phi$ and all the vectors $\xi_k$ in $\Lambda_\psi^m$ about an axis orthogonal to the plane containing $\phi$ and $\psi$ through angle $\theta$; the norm of the inner product of $\psi$ with $\phi$ will increase monatomically from 0 to $\|\phi\|^2$, whilst the norm of the inner product of $\psi$ with each $\xi_k$ will decrease monotonically from $\|\xi_k\|^2 = \|\phi\|^2$ to 0. It follows there is a unique value of $\theta$ where the two are equal, whereupon up to an overall constant, the states $\phi, \xi_k, k = 1, \ldots, m$, rotated through $\theta$ define a new equiamplitude expansion $\Lambda_\psi^{m+1}$ of $\psi$. Since $\Lambda_\psi^2$ can be constructed for any $\psi$, the proof follows by induction. There are uncountably-many, when $\dim \mathcal{H} > 2$, because any equiamplitude expansion of $\psi$ can be rotated about $\psi$ by any angle.

The point of such expansions is to exploit the theorem just proved: any notion of probability satisfying our postulates, that applies to states, must agree that the equiamplitude states in Eq.(6) are equiprobable. But it remains to provide such a notion.

Frequentism in philosophy of probability in the simplest terms is the notion that the probability of $X$ is the fraction of an ensemble that is $X$. Thus, the probability that a card in an ordinary pack of playing cards is an Ace, is the fraction of cards in the pack that are Aces. In particular, each card in the pack occurs just once, so the fraction of each card is the same; if fractions are probabilities, they must be equiprobable, and so counted with equal weight. Natanson (1911) made this point about the use of combinatorial formula, based on frequentism, in the writings of Boltzmann and Planck. According to Jammer (1966 p.51), this was an important step in untangling the riddle of the quantum where it first arose, in Planck's theory of black-body radiation. See also Saunders (2020).

The frequentist probability of the elements of the very ensemble that define probability must all be equal; this is part and parcel of this concept of probability. Yet it has proved to be exceptionally hard to square it with everyday ensembles, usually encountered sequentially -- the probability that the last card in the pack is an Ace is either 0 or 1. Equally obdurate, in everyday cases, is the notion of randomness or equiprobability. However similar the cards in the pack, it is all a matter of how they are shuffled, and how they are dealt.

For these and other reasons, frequentism in philosophy of probability has been mostly abandoned (see, for example, La Caze (2016)), but with little consideration for statistical mechanics -- let alone no-collapse quantum mechanics. When the ensembles are composed of quantum microstates, summing in a superposition to give the physical state, there is no possibility of encountering their elements sequentially, or of shuffling or skewing the superposition, or cheating in which microstate to select: they are all selected, by a unitary evolution, some with $P$ and the rest without, each equally weighted, so the fraction with $P$ is the probability of $P$. If justification is wanted as to why those microstates should be equally weighted, see Section 3.

The other main contrast with classical frequentism is that classically the ensemble could only be spread out in space and in time. In terms of experimental outcomes, this could only mean multiple repetitions of an experiment at different places and times, possibly with slight variations from one run of an experiment to the next. The latter complication afflicts the very definition of classical frequentist probabilities, and likewise their measurement. In the quantum case, the ensemble is defined for a single experimental outcome at a single place and time, with the ensemble made up of vectors in a superposition – the multiplicity is spread out in a superposition.

Of course, to obtain observable statistics, experiments have to be run many times – with those slight variations from one run to the next – but now the complication only concerns the measurement of chance, not its very definition. All this is essential to probability in the epistemic sense, as based on evidence, but that is not our business here.

There is the further question of how one experimental outcome can exist in a superposition with other, different outcomes, and what happens next – we have already remarked on this, it is the Everett interpretation. We are adding an interpretation, not of macroscopic superpositions in terms of worlds changing over time, but of microscopic superpositions in terms of physical probability at each instant of time.

If now there are sentient observers before a measurement, there will be observers in each branch following. The question of self-location arises – in which branch am I located, the branch with spin-up or the branch with spin-down? There is a place for uncertainty, and even a kind of ignorance, albeit of a special sort ('self-locating uncertainty'), a question much discussed in the literature on the Everett interpretation (it was first introduced in Vaidman (1998)). It is arguably present even before the measurement (as observers, prior to measurement, know that that this self-location question is coming, however quickly they may learn the answer). There is a place for postulates in decision theory (as in Wallace (2012); for critical discussion, see Saunders et al (2010)). Again, all these things come with the notion of localised sentient observers, with first-person concerns and limitations, and they may well be essential to making sense of epistemic probability in no-collapse theory; but to the physical, not at all.

## 5. Fractions and limits of fractions as physical probabilities

For the simplest example, consider the special case of projectors $P$ on $\mathcal{H}$ for which expansions $\Lambda_\psi^n$ exist which diagonalise $P$, meaning either that every microstate in $\Lambda_\psi^n$ is an eigenstate of $P$ (the usual notion), or that superpositions of microstates in $\Lambda_\psi^n$ are eigenstates of $P$, leaving no residue.

Let the superposition of $m$ microstates lie in $P$, and let the superposition of the remaining $n - m$ lie in not-$P$: then by microstate-counting, the probability of $P$ is $m/n$. In terms of states: let the superposition of $m$ microstates equal $P\psi$, and of the remaining $n - m$ equal $(I - P)\psi$: then the probability of $P\psi$ is $m/n$.

Under the stated conditions, the Born rule quantity for $P$ in $\psi$ is:

$$\frac{\|P\psi\|^2}{\|\psi\|^2} = \frac{\|P(\xi_1 + \ldots + \xi_n)\|^2}{\|\xi_1 + \ldots + \xi_n\|^2} = \frac{\|(\xi_1 + \ldots + \xi_m)\|^2}{\|\xi_1 + \ldots + \xi_n\|^2} = \frac{m}{n}.$$

The same would obtain for any other equiamplitude expansion $\Lambda_\psi^{n'}$ in which $P$ is diagonal, therefore (because the LHS is the same) $m'/n' = m/n$. Where such a fraction is defined, it is unique.

We should verify that this notion of probability satisfies our postulates, in this special case. Consider the lattice generated by $P\psi$, $(I - P)\psi$; then Eq.(2) is clearly satisfied. For the locality condition (3), $\mu_{\psi_\Sigma}[P]$ is shorthand for $\mu_{\psi_\Sigma}[P\psi]$, likewise $\mu_{U\psi_\Sigma}[P]$ for $\mu_{U\psi_\Sigma}[PU\psi]$. We want the vector to which we are assigning a probability to be exactly the same, on the RHS of Eq.(3), namely $P\psi = \varphi$ (in our previous notation); so we require both $UP\psi = P\psi$ and $PU\psi = P\psi$. Therefore the condition is:

$$UP\psi = P\psi \text{ and } PU\psi = P\psi \Rightarrow \mu_{U\psi_\Sigma}[P] = \mu_{\psi_\Sigma}[P]. \qquad (7)$$

We suppose as before that $P$ is diagonalised in the equiamplitude expansion

$$\psi = \xi_1 + \cdots + \xi_m + \cdots + \xi_{m+1} + \cdots + \xi_n \qquad (8)$$

where $P(\xi_1 + \cdots + \xi_m) = \xi_1 + \cdots + \xi_m$, and $P(\xi_{m+1} + \cdots + \xi_n) = 0$. Consider the expansion

$$U\psi = U(\xi_1 + \cdots + \xi_m) + U(\xi_{m+1} + \cdots + \xi_n) \qquad (9)$$

Where, by hypothesis, $UP\psi = P\psi$, and so conclude
$$U(\xi_1 + \cdots + \xi_m) = \xi_1 + \cdots + \xi_m.$$

Likewise, by hypothesis, $PU\psi = P\psi$, and so conclude
$$PU(\xi_{m+1} + \cdots + \xi_n) = 0.$$

Therefore the fraction of microstates in (8) that lie in $P$, is the same as in (9), so the probability of $P$ is unchanged. From uniqueness, the same will be true in any other equiamplitude expansion of $U\psi$ that diagonalises $P$.

The argument extends easily to include any commuting family of projectors $P$ that can be simultaneously diagonalised in some equiamplitude expansion of $\psi$, but it extends as well to a much larger class – the class of projectors of infinite dimension, whose completements also have infinite dimension (recall from the outset we supposed dim $\mathcal{H} = \infty$). Localisation in space is a simple example; however small the region of space, the associated projector is infinite-dimensional, as is its complement.

This case certainly includes the projectors of interest for Alice and Bob, in the EPRB set-up. It is true we usually concentrate only on spin degrees of freedom, ignoring, for example, the centre of mass degrees of freedom of spin systems – which require an infinite-dimensional Hilbert space – and work with $P$ as a projector on $\mathbb{C}^N$; but really we mean $P \otimes I$; it and its complement $(I - P) \otimes I$ are both infinite-dimensional. Whilst it remains true that in any equiamplitude expansion for $P$ in $\psi$ (meaning $P \otimes I$), in general not all microstates can be eigenstates of $P$, but in practise (since $n$ can be as large as desired) we can ignore the complication, and we will continue to speak of expansions $\Lambda_\psi^n$ that diagonalise $P$, as always available, even though containing at least one Schrödinger-cat state for $P$.

For the details, I refer to Saunders (2024). That treated the case where the continuous degree of freedom enters in the definition of a decoherent history space. There, the limit $n \to \infty$ was not taken, to ensure that microstates were all also decohering states, but as was also there noted (2024 §8), the restriction may well be lifted. In this way probabilities as real numbers (and not just rational numbers) can be derived in the limit $n \to \infty$. (Since identical, in that limit, to the Born rule quantities, we have another way of seeing that P4 must be satisfied: for it is formally satisfied by the Born rule, with no collapse, where $X$ is itself a component of the state.)

This method for defining real number-valued probabilities, and not just rational numbers, is typical in frequentism, classical or quantum, but classically, $n$ is the number of repetitions of an experiment, so the infinite limit was always a fantasy. It was very far from being even a moderately large number (like the number of stars in the visible universe) - the idea of achieving that finite limit was also a fantasy. But in the quantum case, the state really can be expanded into that number of microstates; and to that number to the power of that number of microstates; and so on ad infinitum. Every such expansion involves the same event, the same physical state, the same projector, the same instant of time.

**6. Locality**

It is worth restating that probability as microstate-counting does satisfy our postulates, P1-P4. The postulates – all of them – *follow* on defining probabilities as fractions. To take P3, Kolmogorov's postulates, we noted, follow from Dutch Book arguments, but they also follow for probabilities as fractions: for a fraction of an ensemble, any ensemble, is a real number, bounded by 0 and 1, and disjoint fractions must be additive. The argument to show P4 is satisfied has already been given (we shall shortly give it again, and in a much simpler form, in the context of the Bell set-up). So in claiming, as we do, that physical probability in no-collapse quantum theory is local, we are not merely restating P4, but showing how it may be true. Our postulates resemble a 'principle theory', in Einstein's sense (as set out in his 1905 paper introducing special relativity), whereas microstate-counting is a 'constructive theory' (in special relativity, as defined by electromagnetism in Minkowski space).

To that end, and in the language introduced by Bell (1964), we suppose the hidden variable, denote $\lambda$, belonging to some space $\Lambda$, determines the probabilities of given outcomes $s, t$ of measurements by Alice and Bob on their two spin-systems. These we suppose can be made in one of two directions, as chosen by Alice, denote $a, a'$, and in one of two directions, as chosen by Bob, denote $b, b'$. The outcomes are either spin-up or spin-down, so $s$ and $t$ range over the same two values, $\{+1, -1\}$. The joint probability is then written $p_{a,b}(s,t|\lambda)$.

In effect, from our point of view, this is to replace our physical state $\psi \in \mathcal{H}$ by the hidden variable $\lambda \in \Lambda$, where the latter was likewise supposed to give the 'complete' description of the spin systems, and even – when instrument settings are chosen – of the entire experimental set-up. (This also suggests a new hidden variable theory, where $\lambda \in \Lambda_\psi^n$ is a microstate; see Saunders (2025).)

The marginal probability for Alice's outcome $s$ is the function:

$$p_{a,b}(s|\lambda) = \sum_{t=\pm 1} p_{a,b}(s,t|\lambda). \tag{10}$$

If we consider the events $X$ and $Y$ in Postulate 4 to be the instrument settings $a, a'$, and $b, b'$, then P4 is the statement that Alice's probability of outcome s, on measuring spin in direction a, as given by $p_{a,b}(s|\lambda)$, cannot be changed by Bob's actions alone, with no other change in Alice's vicinity (and similarly vice versa, interchanging the roles of Alice and Bob). When Bob changes his choice from $b$ to $b'$, we conclude from P4:

$$p_{a,b}(s|\lambda) = p_{a,b'}(s|\lambda). \tag{11}$$

But this is just parameter independence (see e.g. Myrvold et al (2024)). It is clearly a locality condition, in that were $\lambda$ controllable, violation of (11) would permit super-luminal signalling. And in our theory, $\lambda$ *is* controllable, for it is the quantum state $\psi \in \mathcal{H}$; so (11) had better not be violated. (It is violated in pilot-wave theory, but there the hidden variable is generally taken to be uncontrollable, and superluminal signalling avoided for that reason.)

We previously showed that quantum fractions as probabilities satisfy P4, in the sense of Eq.(3). That they satisfy parameter independence can be seen much more simply on introducing a tensor product structure for the EPRB set-up that reflects the dynamical independence of Alice from Bob. The probability $p_{a,b}(s,t|\lambda)$ is then the probability of $P_s^a \otimes P_t^b$ in the state $\psi$; the marginal probability for Alice given by (10) is the probability of $P_s^a \otimes I$, and it is given by the fraction of microstates in any equiamplitude expansion that diagonalises $P_s^a$, regardless of Bob's choice of instrument settings – for Bob can only perform local unitary changes of the form $I \otimes U$. Parameter independence (11) is obviously satisfied. And that it is satisfied by the Born rule is equally obvious (this is the no-signalling theorem, true also in ordinary quantum mechanics).

Consider now what P4 requires when the event that changes remotely, the event $Y$, is not Bob's instrument setting, but the *outcome* of his measurement, either spin-up or spin-down. But now we must analyse this through the concept of *change*, that enters P4 – and specifically, change as restricted to purely unitary transformations. What change of this sort can take place in Bob's vicinity, when he performs his measurement, if there is no collapse? The answer is that it can only be the *development of a superposition*. So, the relevant locality condition is that when Bob brings this about – when he not only chooses his instrument settings, but actually performs the measurement, producing the superposition – this should make no difference to Alice's probabilities. But that involves a simple extension of parameter independence, already satisfied. (This point has been made many times before, for example, in Wallace (2012 p.310).)

Similar reasoning applies to Einstein's photon-box thought experiment, which resembles our envelopes and dollar bill: the photon is a superposition, entangled with two boxes, spatially-far apart. In no-collapse theory, when Bob opens his box he simply enters into a superposition, without changing Alice's box or the probability of her finding a photon therein. (For further discussion in no-collapse theory, see Vaidman (2025).)

There is no further requirement, based on locality (so following from P4), in no-collapse quantum theory, to supplement parameter independence. But then it follows that Bell inequalities cannot be derived, as parameter independence is not enough to give factorizability. And the fact that the inequalities can be violated, in no-collapse quantum mechanics, has nothing to do with action-at-a-distance.

Their violation, however, surely shows something. To see what it is, consider the other condition that is often imposed as a locality requirement, namely 'completeness' (often also called 'outcome independence'), which with parameter independence implies factorizability. It is the condition:

$$p_{a,b}(s,t|\lambda) = p_{a,b}(s|\lambda)p_{a,b}(t|\lambda). \tag{12}$$

This clearly is *not* satisfied, in general, when $\lambda$ is the quantum state; but it is when $\psi$ is a product state.

This point is also well-known, but it is worth demonstrating in terms of microstate-counting. Let $\psi = \phi \otimes \chi$. Consider the probability of $P_s^a \otimes P_t^b$ in $\psi$. We expand $\phi$ and $\chi$ in equiamplitude microstates that diagonalise $P_s^a$, $P_t^b$ respectively, of number $n_a$ and $n_b$; of these, let $m_a$ have eigenvalue +1 for $P_s^a$, and let $m_a$ have eigenvalue +1 for $P_t^b$, with all the rest 0 eigenvalues. Then our equiamplitude expansion for $P_s^a \otimes P_t^b$ is the product of $\phi$ and $\chi$ where (on choosing a convenient ordering):

$$\phi = \phi_1^1 + \cdots + \phi_{m_a}^1 + \phi_{m_a+1}^0 + \cdots + \phi_{n_a}^0 \tag{13}$$

$$\chi = \chi_1^1 + \cdots + \chi_{m_b}^1 + \chi_{m_b+1}^0 + \cdots + \chi_{n_b}^0. \tag{14}$$

with superscripts 1 and 0 for eigenstates +1 and 0 respectively. The product $\phi \otimes \chi$ then consists of $n_a \cdot n_b$ equiamplitude microstates each of the form (neglecting, as usual, all degrees of freedom but spin):

$$\phi_j^1 \otimes \chi_k^1;\ \phi_j^0 \otimes \chi_k^1;\ \phi_j^1 \otimes \chi_k^0;\ \phi_j^0 \otimes \chi_k^0\ . \tag{15}$$

By construction, $m_a \cdot m_b$ of these are +1 eigenstates of $P_s^a \otimes P_t^b$. So, the probability of $P_s^a \otimes P_t^b$ is

$$p_{a,b}(s,t|\psi) = \frac{m_a \cdot m_b}{n_a \cdot n_b}.$$

Similarly, the probabilities of $P_s^a \otimes I$ and $I \otimes P_t^b$ are respectively

$$p_{a,b}(s|\psi) = \frac{m_a}{n_a},\ p_{a,b}(t|\psi) = \frac{m_b}{n_b}$$

and the completeness condition is automatically satisfied.

This reasoning breaks down when $\psi$ is not a product state – when it is entangled, with spin degrees of freedom in the singlet state of spin. The microstates will still be an expansion of terms of the form Eq.(15), if Alice and Bob measure spin in directions $a$, $b$, respectively; but they can no longer be obtained as the product of two expansions (13), (14). But this is to just restate the fact that it is an entanglement. The point has nothing to do with non-locality.

To understand how (12) does arise as a locality condition, we need a single-world perspective, in which, when Bob performs this remote experiment, whether by collapse or by any other means, a unique outcome is produced. Embedded in any theory like that, that outcome *is* the change in $Y$, and by P4, ought to leave the probabilities for Alice's outcome unchanged. That suggests the condition (along with its conjugate)

$$p_{a,b}(s|\lambda) = p_{a,b}(s|t,\lambda)\ . \tag{16}$$

This condition is known as 'conditional outcome independence'; it implies completeness (12). But it appears to depend essentially on the idea of uniqueness of outcome. If unique, then $Y$ is changed, not into a superposition of two outcomes, but into a single outcome. By P4, the probability of $X$ should not be changed, implying (16). Its violation *does* imply action-at-a-distance. But if the change in $Y$ is the development of a superposition, that is already looked after by parameter independence.

Also produced are various correlations between Alice's outcomes, and Bob's. There are several such, and they are all relations. But relations can be changed by changes in the relata, singly or jointly, without any action-at-a-distance, as witness spatial relations; on those P4 poses no constraint. (Brown and Timpson (2016 §9.2) make this point.)

If this reasoning is correct – if in a single world conditional outcome independence is indeed an expression of locality (and we are agreed that parameter independence is) -- then on the premise of a single world, and absent retrocausation and the like (on this point see Saunders 2025), the observed violations of Bell inequalities imply non-locality. But if that premise is in question, there is the alternative reading: those experiments, together with locality, imply many worlds.

I suggest this is their true meaning, and that never was a Nobel prize for physics, as won by Aspect, Clauser, and Zeilinger in 2022, of greater import.

**Acknowledgments**

My thanks to Hans-Thomas Elze for his encouragement and hospitality at the DICE conferences 2022 and 2024. I am grateful to Charles Bédard, Harvey Brown, Jeremy Butterfield, Cristi Stoica, and David Wallace, for careful readings, comments, and criticisms. Any mistakes that remain are my own.